# The Superior Knowledge Proximity Measure for Patent Mapping

**Bowen Yan**

SUTD-MIT International Design Centre
Singapore University of Technology and Design
8 Somapah Road, Singapore 487372
Email: bowen_yan@sutd.edu.sg

**Jianxi Luo**

Engineering Product Development Pillar & SUTD-MIT International Design Centre
Singapore University of Technology and Design
8 Somapah Road, Singapore 487372
Email: luo@sutd.edu.sg

**Abstract**

Network maps of patent classes have been widely used to analyze the coherence and diversification of technology or knowledge positions of inventors, firms, industries, regions, and so on. To create such networks, a measure is required to associate different classes of patents in the patent database and often indicates knowledge proximity (or distance). Prior studies have used a variety of knowledge proximity measures based on different perspectives and association rules. It is unclear how to consistently assess and compare them, and which ones are superior for constructing a generally useful total patent class network. Such uncertainty has limited the generality and applications of the previously reported maps. Herein, we use a statistical method to identify the superior proximity measure from a comprehensive set of typical measures, by evaluating and comparing their explanatory powers on the historical expansions of the patent portfolios of individual inventors and organizations across different patent classes. Based on the complete United States granted patent database from 1976 to 2017, our analysis identifies a reference-based Jaccard index as the statistically superior measure, for explaining the historical diversifications and predicting future movement directions of both individual inventors and organizations across technology domains.

# 1. Introduction

Recent studies in the information science literature have presented various network maps of patent classes (Alstott et al., 2017b; Engelsman & van Raan, 1994; Joo & Kim, 2010; Kay et al., 2014; Leydesdorff et al., 2014; Nakamura et al., 2015; Yan & Luo, 2017b). In such a network, the vertices are patent classes in a patent classification system and approximate technology fields (e.g., organic chemistry), and the edges between the vertices are weighted according to the knowledge proximity (or distance) between technology fields and measured based on massive patent document information. Such network maps cover all technology classes in the patenting system and utilize the entire patent database to compute knowledge proximity deriving statistical significance, and thus are considered representations of the total technology space (Alstott et al., 2017b). Figure 1 is an example of the patent technology network map.

Such patent technology networks have been used to analyze the patent portfolio diversification of regions (Boschma et al., 2014; Rigby, 2015), firms (Breschi et al., 2003; Luo et al., 2017; Teece et al., 1994), individuals (Alstott et al., 2017a) and system products (Song et al., 2016) across different technology fields. These studies consistently showed that the diversification, regardless of levels of analysis, is more likely into more proximate fields than distant ones in the technology space and in turn, suggest the predictive power of knowledge proximity on technology diversifications and the search for innovations (Frenken et al., 2007; Leten et al., 2007). However, these studies used various measures of knowledge proximity based on different perspectives and association rules. It is unclear which measures are generally superior for constructing the total technology network maps. This ambiguity has limited the use of technology network maps for general purposes and contexts. We address this challenge in this research.

Herein we use a statistical method to evaluate and compare the explanatory powers of alternative knowledge proximity metrics on the historical expansions of the patent portfolios of individual inventors, organizations, regions or design domains across different patent classes. By statistically comparing various metrics using the USPTO data, the reference-based Jaccard index that measures the extent of knowledge base overlapping provides the highest explanatory power on the next domains of individual inventors and organizations given their previous domains. In turn, this metric and its resulting technology space map are recommended for prescriptive and predictive analyses in general contexts. Figure 1 illustrates an example of the total technology space map built using the superior measure on the 3-digit CPCs.

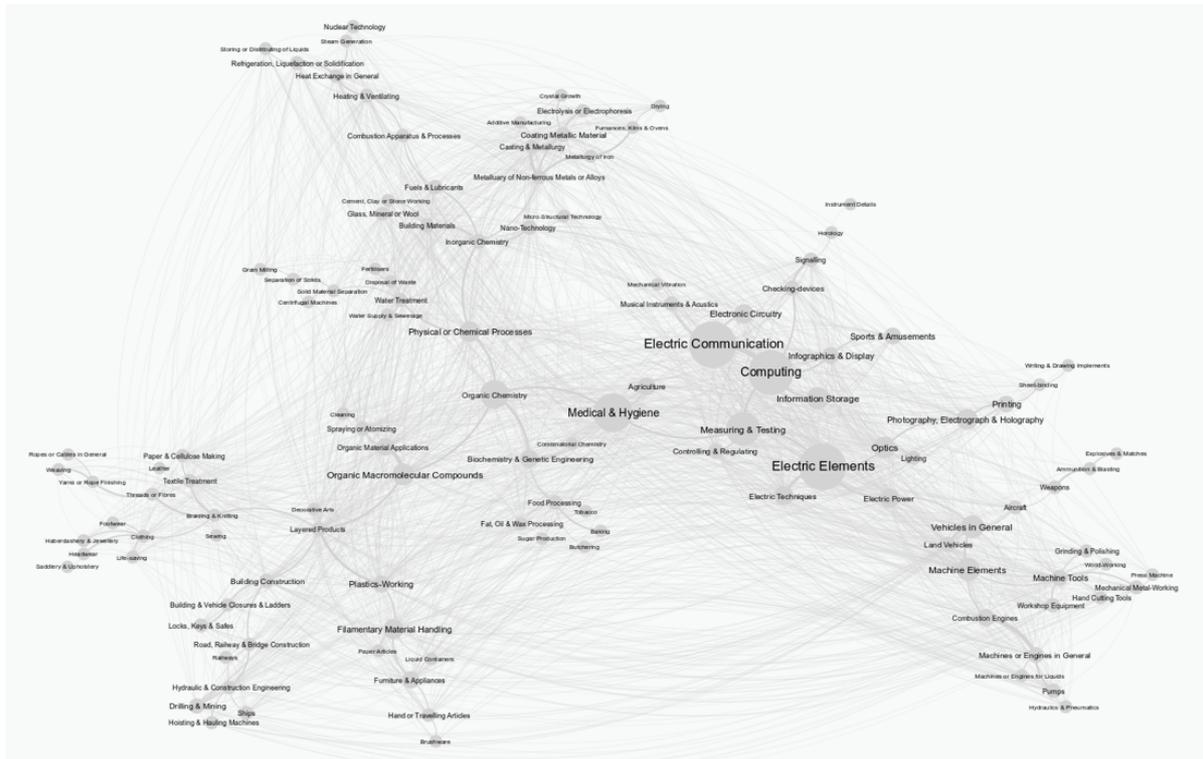

Figure 1. The total technology space map using class-to-patent Jaccard based on references to measure knowledge proximity between the 122 3-digit patent classes in CPC. The original network is extremely dense, so the visualization here only contains the maximum spanning tree (i.e., the minimal set of 121 edges that connect all vertices and maximize total edge weights) as the backbone plus additional 1,069 strongest edges from the original network. The network filtering threshhold was determiend using a technique from Yan and Luo (2017a) that aims to remove as many weak edges as possible but maintain as much explanatory power of the network as possible.

## 2. Literature review

### 2.1 Knowledge Proximity and Diversification

Prior empirical studies have shown statistical evidence that firms and regions are more likely to diversify across technology areas with high knowledge proximity, because of the ease to master new but proximate knowledge. For instance, at the city level, Rigby (2015) showed that U.S. cities' entries into and exits from technology domains are highly related to the knowledge proximity among cities' prior and next technology domains. He measured knowledge proximity as the probability that a patent in one class will cite a patent in the other. At the firm level, Breschi et al. (2003) found firms in Europe are more likely to diversify across technology domains with high knowledge proximity. They used the patents from European Patent Office and patent co-classification codes to measure knowledge proximity between technology fields. Luo et al. (2017) used the technology space map to analyze the evolution of technology capability positions of Google with regard to its driverless car project, and found Google followed the strongest network paths to grow its technology positions over time in the total technology space. They used the cosine similarity between the class-to-patent citation vectors to calculate the knowledge proximity.

At the individual level, an analysis of 2 million inventors in 4 million USPTO patents showed that inventors are much more likely to explore technology domains that are more proximity to their prior patenting domains (Alstott et al., 2017a). They used a normalized knowledge proximity metric by comparing direct empirical patent citations from one domain to another to the same parameter in randomized patent citation networks. Srinivasan et al. (2018) used a technology space map to gauge the impact of knowledge distance on design creativity based on a human experiment. They found that engineers are more likely to identify inspirational and useful patents in the domains near their home domains for concept generation, but more novel concepts are inspired by those patents from more distant domains. Their map is based on the Jaccard index of inter-class referencing vectors. Luo et al. (2018) further proposed to use a total technology space map as a heuristic ideation tool to enhance design opportunity conception. They used a cosine similarity metric to calculate knowledge proximity and create the map and demonstrate its uses for ideation in a few human experiments.

In sum, these prior works have consistently suggested the explanatory power of knowledge proximity, despite a variety of knowledge proximity metrics, on the expansions of the patent portfolios of inventors, firms, regions or design domains as a subgraph of the total space map, and the utility of the patent technology network to prescribe the directions of innovation and diversification of inventors and firms. However, we are still faced with the uncertainty in the selection of knowledge proximity metrics, given the existence of many alternatives.

*2.2 Knowledge Proximity Measures*

There are two major groups of patent data-based measures of knowledge proximity in the literature. One group of measures uses patent reference information. For instance, the co-reference measure takes the form of Jaccard index (Jaccard, 1901; Yan & Luo, 2017b) to calculate the count of shared references of pairs of classes normalized by the total count of all unique patent references in either class (Iwan von Wartburg et al., 2005; Leydesdorff & Vaughan, 2006). Leydesdorff et al. (2014) and Kay et al. (2014) used the cosine similarity measure, i.e., the cosine of the vectors of patent references made from a pair of classes to all other classes respectively. For a higher granularity, Yan and Luo (2017b) applied the cosine similarity measure to class-to-patent vectors, concerning references to specific patents instead of aggregated classes.

Another group of measures mines the patent co-classification information, i.e., how often two classes are co-assigned to the same patents (Engelsman & van Raan, 1994). Using this information, the relatedness between patent classes can be measured according to the co-occurrence of classification codes assigned to patent documents (Engelsman & van Raan, 1994). The assumption is that the frequency in which two classes are jointly assigned to the same patents infers the knowledge proximity of the classes. For example, Breschi et al. (2003) measured knowledge proximity between patent classes as the cosine of respective patent classes' vectors of occurrences with all other classes in patents. Nesta and Dibiaggio (2005) measured the deviation of the actual observed co-occurrences of class pairs in patents from random expectations.

Furthermore, prior studies (Hinze et al., 1997; Yan & Luo, 2017b) have shown that the structures of the patent class networks are consistent over time, regardless of the choices of knowledge proximity measures to create the networks. For example, Yan and Luo (2017b)'s longitudinal analysis showed that the changes of all links' weights and their relative rankings by weights over different years and decades are small and insignificant. Such stability of the measurements may be the result of the innate but latent proximity or distance between different physical technologies, e.g., computing and combustion engine. That is, the proximity or distance between physical technologies has an innate physical nature; the technology space is also a latent physical existence. Therefore, the approximations of the technology space using data from different time periods are not supposed to vary, if sufficient data are computed in a chosen time period and statistical significance is ensured. In turn, the stability of the patent technology networks has allowed them for the analysis of over-time patent portfolio diversification of individuals, firms, regions and system products.

## 3. Methodology
*3.1 Data*

We construct the total technology space networks and analyze the historical patent portfolios using the granted utility patent data of the United States Patent and Trademark Office (USPTO). The data set contains more than 6 million utility patents that can be downloaded from PatentsView[1]. Each patent is categorized using one or more Cooperative Patent Classification (CPC) codes (i.e., patent classes), which have been jointly developed by the European Patent Office (EPO) and the USPTO. The CPC system includes 9 sections, which are subdivided into classes, subclasses, main groups, and subgroups. In this study, we use 127 3-digit level classes and 654 4-digit level subclasses as the vertices in the patent class networks. PatentsView's data sources provide disambiguated inventor and assignee identifiers, which distinguish whether inventors or assignees with the same name are the same person or the same assignee. We use the unique identifiers to build historical patent portfolios for inventors and assignees.

---

[1] Data is available at http://www.patentsview.org/download/.

3.2 Knowledge proximity measures

To evaluate the relationships among patent classes, the metrics in the literature have mainly taken two mathematical forms: *Jaccard index* and *cosine similarity*. Equations (1) and (2) are the general formulas for calculating Jaccard index and cosine similarity between any pair of patent classes.

$$Jaccard\ index = \frac{|C_i \cap C_j|}{|C_i \cup C_j|} \tag{1}$$

$$Cosine\ similarity = \frac{\sum_k C_{ik} C_{jk}}{\sqrt{\sum_k C_{ik}^2} \sqrt{\sum_k C_{jk}^2}} \tag{2}$$

The variables in the equations can be operationalized differently according to the information in the patent documents to calculate these metrics. In the following, we list the Jaccard index and cosine similarity metrics based on either reference or co-classification information.

A1. *Jaccard class-to-patent based on reference*. It is the count of shared patent references, normalized by the total count of all unique references of patents in a pair of patent classes. With equation (1), $C_i$ and $C_j$ are the sets of the references of the patents in patent classes $i$ and $j$; $|C_i \cap C_j|$ is the number of patents referenced in both patent classes $i$ and $j$, and $|C_i \cup C_j|$ is the total number of unique patents referenced in both patent classes $i$ and $j$, respectively.

A2. *Jaccard class-to-class based on reference*. It is the count of shared patent classes that were assigned to references of patents, normalized by the total count of all unique patent classes assigned to the reference of the patents in a pair of patent classes. With equation (1), $C_i$ and $C_j$ are the sets of patent classes assigned to the references of patents in patent classes $i$ and $j$, respectively; $|C_i \cap C_j|$ is the number of patent classes assigned to references of patents in both patent classes $i$ and $j$, and $|C_i \cup C_j|$ is the total number of unique patent classes assigned to references in both patent classes $i$ and $j$, respectively.

A3. *Jaccard class-to-patent based on co-classification*. It is the count of shared patents, normalized by the total count of all unique patents in the pair of patent classes. With equation

(1), $C_i$ and $C_j$ are the sets of patents that were assigned to patent classes $i$ and $j$; $|C_i \cap C_j|$ is the number of patents assigned to both patent classes $i$ and $j$, and $|C_i \cup C_j|$ is the number of unique patents assigned to either patent classes $i$ or $j$.

A4. *Jaccard class-to-class based on co-classification*. It is the count of shared co-occurring patent classes on the same patents, normalized by the total count of all unique co-occurring patent classes of a pair of patent classes on the same patents. With equation (1), $C_i$ and $C_j$ are the sets of co-occurring patent classes of patent classes $i$ and $j$ on the same patents; $|C_i \cap C_j|$ is the number of co-occurring patent classes of both patent classes $i$ and $j$, and $|C_i \cup C_j|$ is the number of unique co-occurring patent classes of either patent classes $i$ or $j$.

B1. *Cosine class-to-patent based on reference*. It measures the angular similarity between the two vectors representing two patent classes' distributions of references into specific unique patents. With equation (2), $C_{ij}$ denotes the number of references of all patents in class $i$ to the specific patent $j$.

B2. *Cosine class-to-class based on reference*. It measures the angular similarity between the two vectors representing two patent classes' distributions of citations into all patent classes. With equation (2), $C_{ij}$ denotes the number of references of all patents in patent class $i$ to all patents in patent class $j$.

B3. *Cosine class-to-patent based on co-classification*. It measures the angular similarity between the two vectors representing two patent classes' distributions of assignments to specific unique patents. With equation (2), $C_{ij}$ denotes the assignment of class $i$ to patent $j$. $C_{ij} = 1$, if class $i$ is assigned to patent $j$; otherwise $C_{ij} = 0$.

B4. *Cosine class-to-class based on co-classification*. It measures the angular similarity between the two vectors representing two patent classes' distributions of co-occurrences with all other patent classes on the same patents. With equation (2), $C_{ij}$ denotes the number of patents assigned to both patent class $i$ and patent class $j$.

*3.3 Finding the optimal knowledge proximity metric*

In the following, we present a statistical method to identify the optimal knowledge proximity measure among alternatives for constructing the patent class networks that approximate the total technology space. The proposed method consists of three steps depicted in Figure 2:

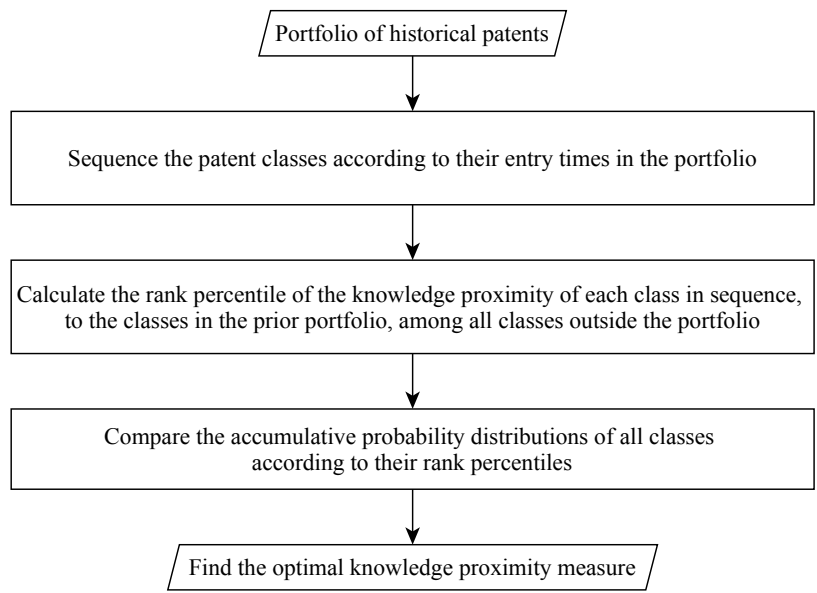

Figure 2. The procedure of finding the optimal knowledge proximity measure.

1) Identify the set of a specific agent's patent classes, according to its historical patent portfolio. The agent can be an individual person, firm, R&D organization, region or a design practice domain (e.g., autonomous vehicle). A patent class is identified to be entered by the agent, if it has patents in that class. And then, we determine the sequence of these patent classes according to the time when the portfolio included patents in a class for the first time. Each following patent class in the sequence is considered as a newly entered class, relative to the previous classes in the sequence. It infers the expansion trajectory of the agent's patent portfolio in the total technology space network. We use *filing date* to determine the time of new class entrances, which is closer to the actual time of inventing than the *grant date*.

2) Consider each patent class (e.g., class D in Figure 3) in the sequence, at its time of being included into the focal portfolio, we calculate the rank percentile of its knowledge proximity to the classes that had been included in the portfolio (e.g., classes A, B, and C in Figure 3) prior to its own inclusion, relative to the proximity of all other unentered classes (e.g., classes E and F in Figure 3) outside the portfolio to those classes in the portfolio, at the time. A higher rank percentile value suggests that the patent class was more strongly related to the classes that had been included in the portfolio, and it was more likely to enter that class. It follows the principle that an agent (e.g., persons, companies, regions, and design practice domains) preferentially expands to new technology fields that are more knowledge-proximate to those which it has already entered. The calculation of the proximity between a class outside the portfolio (e.g., class D in Figure 3) and the classes in the portfolio (e.g., classes A, B, and C in Figure 3) can be executed in various ways. In the analysis later, we use the sum of proximity values of all edges between the outside class and those classes in the portfolio.

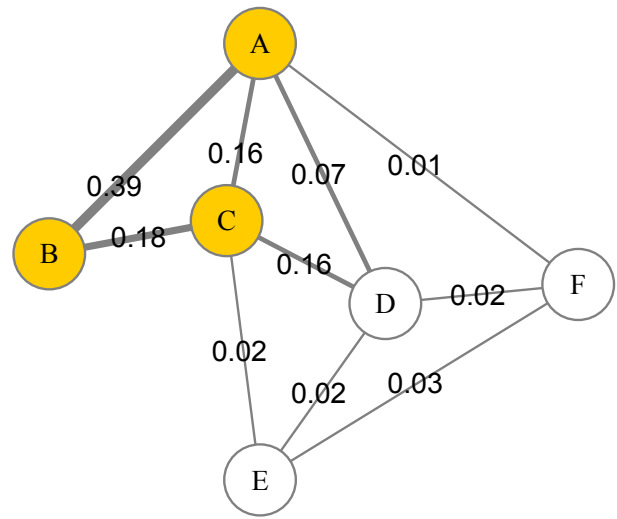

Figure 3. Different types of classes in or outside the sequence. Edge width corresponds to knowledge proximity. Classes A, B, and C are the classes that have been entered. Classes D, E and F are the classes that may be entered next. In this example, the knowledge proximities of classes D, E and F to the classes A, B, and C in the portfolio are 0.23, 0.02, and 0.01, respectively. The rank percentiles of classes D, E, and F are 1, 0.67, and 0.33. Class D is most likely to enter the portfolio and join classes A, B and C.

3) Use different knowledge proximity measures to plot the cumulative portfolio entry probability distributions of classes by their proximity percentile values. Figure 4 illustrates the distributions using different knowledge proximity measures. The dashed line represents the probability given by the null hypothesis: new patent classes are entered randomly regardless of their knowledge proximity values to the previously entered classes of the agent. Therefore, when the empirical curve is above the dashed line, one can say that the agent historical exploration of new classes is conditioned by their proximity values to its prior classes. The knowledge proximity measure that yields the steepest curve (or the closest curve to 100%) and thus the patent class network based on it will provide relatively the highest explanatory power on the agent's cross-domain moves in the total technology space, or on the expansion directions of the patent portfolio in the network of all patent classes. For example, Figure 4 shows measure #3 provides the highest explanatory power, while measure #1 is the lest explanatory.

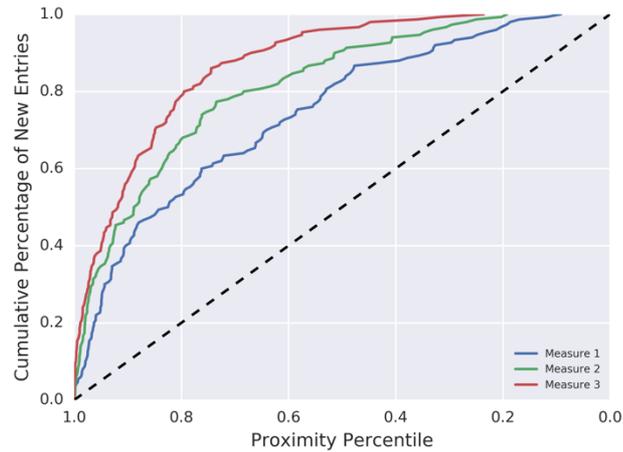

Figure 4. Cumulative distributions of the proximity percentiles of newly entered classes of a specific agent based on different knowledge proximity measures.

In turn, the knowledge proximity measure that provides the highest explanatory power based on a historical patent portfolio can be utilized for predictive analysis of the patent portfolio's future evolution as well as for prescriptive recommendations of new domains for an agent to consider for exploration.

## 4. Results

First, we choose the 3-digit level (CPC3) and 4-digit level (CPC4) patent classes to represent vertices in patent class networks, and edges between pairs of them are weighted by the alternative knowledge proximity measures given in Section 3.2. Based on these CPC3 and CPC4 patent class networks, we measure the proximity percentiles for the historical patent portfolios of more than 120,000 inventors and more than 25,000 assignees who had patented in at least 10 CPC3 and CPC4 classes, respectively, to ensure statistical significance. Figure 5 shows the cumulative distributions of the proximity percentiles of newly entered classes of all selected *inventors* based on CPC3 and CPC4 patent class networks, respectively. Figure 6 shows the cumulative distributions of the proximity percentiles of newly entered classes of all selected *assignees* based on CPC3 and CPC4 patent class networks, respectively.

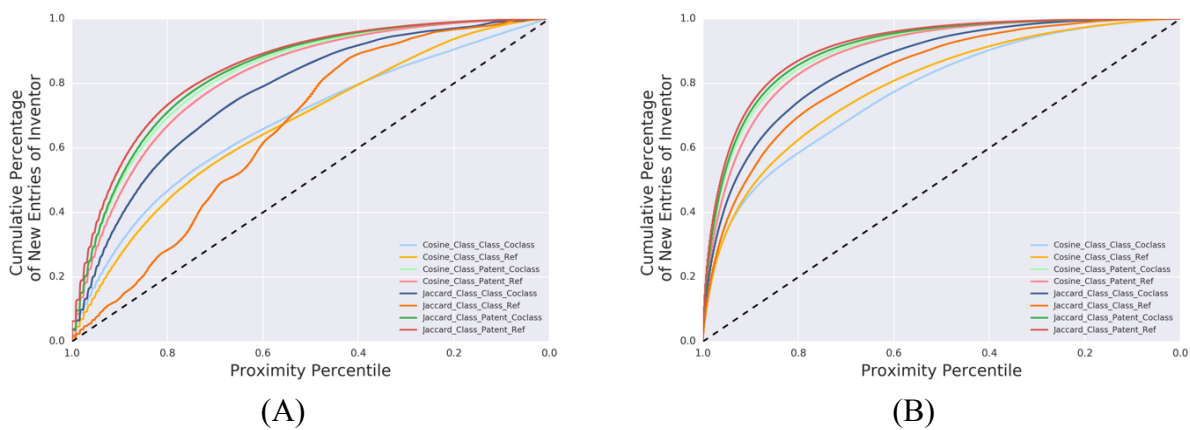

Figure 5. Cumulative distributions of the proximity percentiles of newly entered classes of inventors based on alternative patent class networks. (A) CPC3 patent class networks; (B) CPC4 patent class networks.

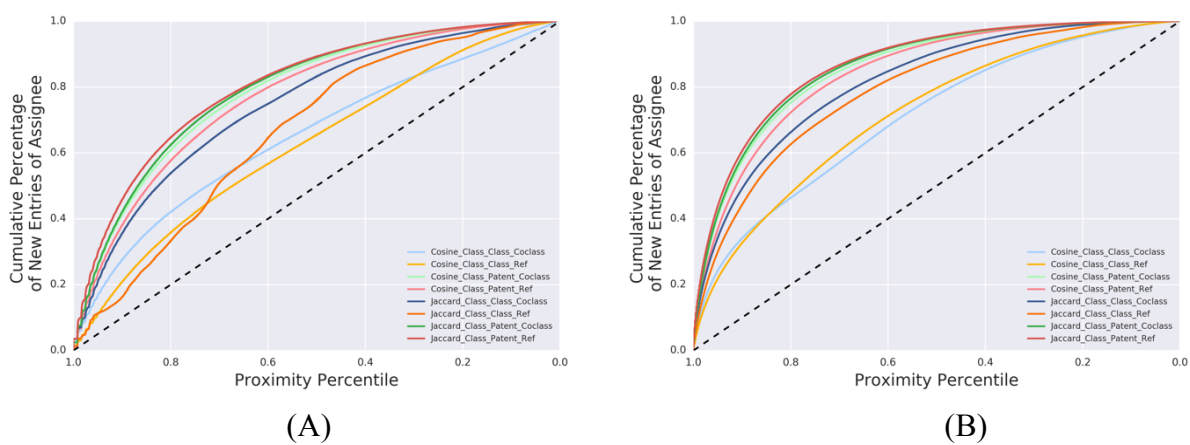

Figure 6. Cumulative distributions of the proximity percentiles of newly entered classes of assignees based on alternative patent class networks. (A) CPC3 patent class networks; (B) CPC4 patent class networks.

Comparing the curves shown in Figures 5 and 6, the *Jaccard class-to-patent based on reference* provides the highest explanatory power on inventor and assignee diversifications, with both CPC3 and CPC4 classes. Regardless of the patent document information (reference or co-classification) used for calculations, the group of measures using the Jaccard index always perform better than the group of measures using the cosine similarity. It is also noteworthy that the CPC4 networks generally provide higher explanatory power than the CPC3 networks, as suggested by larger areas underneath the curves in Figures 5(B) and 6(B) than the areas underneath the corresponding curves in Figures 5A and 6A. In Figure 5(B), using the

*Jaccard class-to-patent based on reference* with CPC4 classes, the top 20% of strongly related patent classes represent about 90% of newly entered classes. In Figure 6(B), using the *Jaccard class-to-patent based on reference* with CPC4 classes, the top 20% of strongly related patent classes represent about 80% of newly entered classes.

We also examine the optimal knowledge proximity measure for every individual inventor and assignee based on their historical patent portfolio. Figure 7 and Figure 8 report the proportions of different knowledge proximity measures selected as the optimal one for each individual inventor or assignee, respectively. *Jaccard class-to-patent based on reference* stands out as the optimal knowledge proximity metric to explain the diversifications of most inventors and assignees in history and is followed by *Jaccard class-to-patent co-classification.* In Figures 7(A) and 8(A), the proportion of *Cosine class-to-patent co-classification* is close to *Jaccard class-to-patent co-classification* method. These results consistently suggest a single superior knowledge proximity measure for creating the total technology space map – *Jaccard class-to-patent based on reference.*

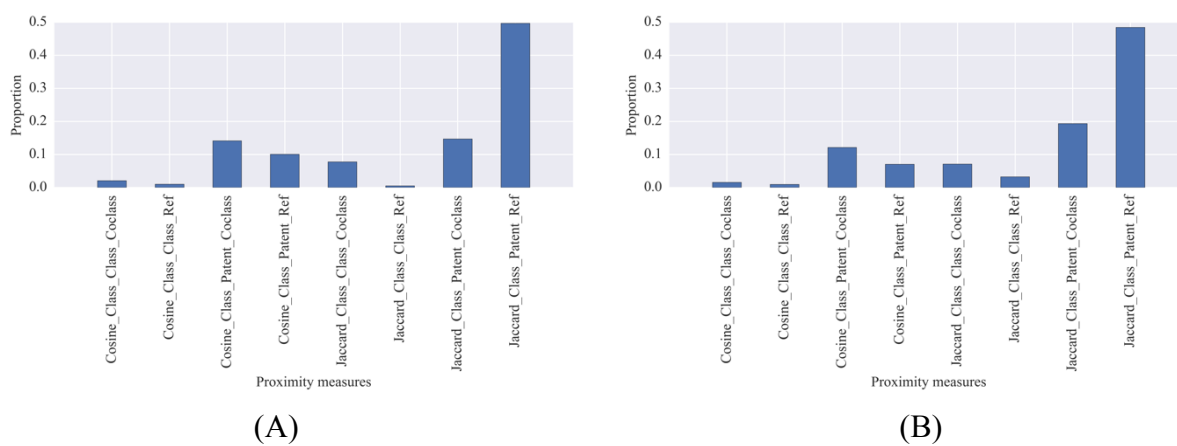

Figure 7. Proportion of different knowledge proximity measures selected as the optimal one for each inventor, using (A) CPC3 classes; (B) CPC4 classes.

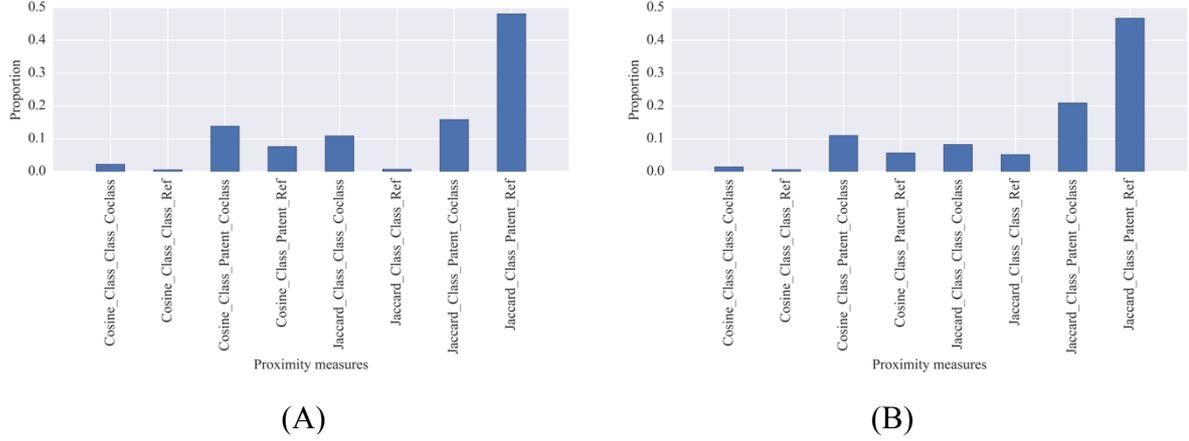

Figure 8. Proportion of different knowledge proximity measures selected as the optimal one for each assignee, using (A) CPC3 classes; (B) CPC4 classes.

## 5. Discussion

In this paper, we mainly compared 8 proximity measures based on either Jaccard index or cosine similarity. In the literature, there are other similarity measures. As a robustness check, we have evaluated the explanatory powers of a few more metrics against the same empirical bases in section 4. For instance, *Pearson's correlation coefficient* is often used as a similarity measure, whereas others have argued it might not be suitable for the cases where the vectors have many zeros (Ahlgren et al. 2003) and have cell values that have already indicated similarity (Leydesdorff & Vaughan, 2006). The *relative entropy* metric from information theory measures the likelihood of information transmission from one system to the other (Cover & Thomas, 2012; Leydesdorff, 1991) and thus may be utilized to indicate knowledge proximity between two patent classes, using the following formula

$$D = \left[\sum_k p_k \log\left(\frac{p_k}{q_k}\right) + \sum_k q_k \log\left(\frac{q_k}{p_k}\right)\right] / 2 \qquad (3)$$

where $p_k$ and $q_k$ represent any two patent classes' probability distributions of references or co-classifications.

Tables 1-4 summarizes the explanatory powers of different measures, including Pearson and entropy, on the same empirical bases in section 4. The size of the area underneath an accumulative distribution curve in Figure 4 is used to quantify the *explanatory power* of the corresponding knowledge proximity metric. *Jaccard class-to-patent based on reference* remains the superior knowledge proximity measure. *Jaccard*-based measures appear to perform better than other measures across different data choices (Reference and Co-classification; Class to Class and to Patent; CPC3 and CPC4).

Table 1. Diversification explanatory power for different measures at the CPC3 level on different data choices for assignees.

|  | Measures of Knowledge Proximity | | | |
|---|---|---|---|---|
|  | Jaccard | Cosine | Pearson | Entropy |
| Class_Class Reference | 0.665 | 0.623 | 0.636 | 0.370 |
| Class_Patent Reference | **0.803** | 0.771 | 0.686 | 0.692 |
| Class_Class Co-Classification | 0.746 | 0.648 | 0.668 | 0.412 |
| Class_Patent Co-Classification | 0.794 | 0.786 | 0.741 | 0.740 |

Table 2. Diversification explanatory power for different measures at the CPC4 level on different data choices for assignees.

|  | Measures of Knowledge Proximity | | | |
|---|---|---|---|---|
|  | Jaccard | Cosine | Pearson | Entropy |
| Class_Class Reference | 0.795 | 0.720 | 0.724 | 0.316 |
| Class_Patent Reference | **0.870** | 0.846 | 0.633 | 0.735 |
| Class_Class Co-Classification | 0.818 | 0.709 | 0.715 | 0.359 |
| Class_Patent Co-Classification | 0.864 | 0.858 | 0.827 | 0.780 |

Table 3. Diversification explanatory power for different measures at the CPC3 level on different data choices for inventors.

|  | Measures of Knowledge Proximity | | | |
|---|---|---|---|---|
|  | Jaccard | Cosine | Pearson | Entropy |
| Class_Class Reference | 0.655 | 0.673 | 0.688 | 0.339 |
| Class_Patent Reference | **0.844** | 0.815 | 0.747 | 0.700 |
| Class_Class Co-Classification | 0.769 | 0.678 | 0.709 | 0.396 |
| Class_Patent Co-Classification | 0.834 | 0.827 | 0.786 | 0.776 |

Table 4. Diversification explanatory power for different measures at the CPC4 level on different data choices for inventors.

|  | Measures of Knowledge Proximity | | | |
|---|---|---|---|---|
|  | Jaccard | Cosine | Pearson | Entropy |
| Class_Class Reference | 0.830 | 0.793 | 0.797 | 0.271 |
| Class_Patent Reference | **0.913** | 0.892 | 0.726 | 0.735 |
| Class_Class Co-Classification | 0.858 | 0.775 | 0.784 | 0.271 |
| Class_Patent Co-Classification | 0.906 | 0.900 | 0.877 | 0.817 |

## 6. Conclusion

In this paper, we have identified a consistently superior knowledge proximity measure, among a set of typical measures, in explaining the historical diversifications of the patent portfolios of individual inventors and assignees in the USPTO patent database. The superior knowledge proximity measure is *Jaccard class-to-patent based on reference*. In an earlier study of (Yan & Luo, 2017b), they found this metric results in patent class map structures that are most correlated with the map structures resulting from other alternative metrics. In addition to this specific measure, we also find that all the measures using the Jaccard index provide better explanation of inventor and assignee diversifications than the measures using the cosine similarity.

The superior knowledge proximity measure is then recommended for constructing the patent class network map to approximate the total technology space, and for applications of analyzing historical and predicting future expansions of movement directions of the technology positions of individual investors and assignees. However, the superior metric is not superior for all individuals and assignees. For instance, in our investigations, the *Jaccard class-to-patent based on co-classification* provides the best explanation of the diversification of Apple Inc when using CPC4 classes; and the *cosine class-to-patent based on reference* presents the highest explanatory power of the diversification of the inventor Shunpei Yamazaki when using CPC3 classes. Therefore, we plan to develop a computer tool that would allow the intelligent identification of the best knowledge proximity metric for the descriptive, predictive and prescriptive analytics of specific inventors or assignees. Furthermore, despite our focus on inventors and assignees, such a tool would be also useful for the analyses of the patent portfolio of a region (e.g., Boston, MA or Huston, TX) or an industry or design practice domain (e.g., 3D printing or autonomous vehicle) in the total technology space.

**Acknowledgement**

We would like to thank Loet Leydesdorff for his useful comments and valuable suggestions. This research is funded by the SUTD-MIT International Design Centre.